\documentclass[aps,prl,epsf,superscriptaddress,amsmath,amssymb,amsfonts,twocolumn,showpacs]{revtex4-1}

\usepackage{graphicx}
\usepackage{epsfig}
\usepackage{dcolumn}
\usepackage{bm}
\usepackage{braket}
\usepackage{amsmath}
\usepackage{color} 
\newcommand{\abs}[1]{\left| #1 \right|}

\usepackage{tcolorbox}
\usepackage{tabularx}
\usepackage{array}
\usepackage{colortbl}
\tcbuselibrary{skins}

\newcolumntype{Y}{>{\raggedleft\arraybackslash}X}

\tcbset{tab1/.style={fonttitle=\bfseries\large,fontupper=\normalsize\sffamily,
colback=yellow!10!white,colframe=red!75!black,colbacktitle=white!40!white,
coltitle=black,center title,freelance,frame code={
\foreach \n in {north east,north west,south east,south west}
{\path [fill=red!75!black] (interior.\n) circle (3mm); };},}}

\tcbset{tab2/.style={enhanced,fonttitle=\bfseries,fontupper=\normalsize\sffamily,
colback=yellow!6!white,colframe=red!50!black,colbacktitle=yellow!20!white,
coltitle=black,center title}}

\begin{document}
\title{Quench Dynamics and Orthogonality Catastrophe of Bose Polarons}

\author{S.I. Mistakidis}
\affiliation{Center for Optical Quantum Technologies, Department of Physics, University of Hamburg, 
Luruper Chaussee 149, 22761 Hamburg Germany}

\author{G.C. Katsimiga}
\affiliation{Center for Optical Quantum Technologies, Department of Physics, University of Hamburg, 
Luruper Chaussee 149, 22761 Hamburg Germany}

\author{G.M. Koutentakis}
\affiliation{Center for Optical Quantum Technologies, Department of Physics, University of Hamburg, 
Luruper Chaussee 149, 22761 Hamburg Germany}\affiliation{The Hamburg Centre for Ultrafast Imaging,
Universit\"{a}t Hamburg, Luruper Chaussee 149, 22761 Hamburg,
Germany}

\author{Th. Busch}
\affiliation{Quantum Systems Unit, OIST Graduate University, Onna, Okinawa 904-0495, Japan}

\author{P. Schmelcher}
\affiliation{Center for Optical Quantum Technologies, Department of Physics, University of Hamburg, 
Luruper Chaussee 149, 22761 Hamburg Germany} \affiliation{The Hamburg Centre for Ultrafast Imaging,
Universit\"{a}t Hamburg, Luruper Chaussee 149, 22761 Hamburg,
Germany}

\date{\today}

\begin{abstract}
We monitor the correlated quench induced dynamical dressing of a spinor 
impurity repulsively interacting with a Bose-Einstein condensate. 
Inspecting the temporal evolution of the structure factor three distinct dynamical regions arise upon increasing the interspecies 
interaction. 
These regions are found to be related to the segregated nature of the impurity and to the ohmic character of the bath. 
It is shown that the impurity dynamics can be described by an effective potential that deforms from a harmonic to a double-well one
when crossing the miscibility-immiscibility threshold. 
In particular, for miscible components the polaron formation is imprinted on the spectral response of the system. 
We further illustrate that for increasing interaction an orthogonality catastrophe occurs 
and the polaron picture breaks down. Then a dissipative motion of the impurity takes place 
leading to a transfer of energy to its environment. This process signals the presence of entanglement 
in the many-body system.
\end{abstract}

\maketitle

{\it Introduction.--}
A valuable asset of ultracold atoms is the opportunity to track the real time dynamics of 
quantum many-body (MB) systems such as multi-component quantum gases composed of 
different atomic species~\cite{Modugno} or different hyperfine states of the same species~\cite{Myatt,Stenger}. 
In particular, the realization of highly population imbalanced atomic gases 
with tunable interactions~\cite{Wu,Heo,Cumby,Roati,Pilch,Schirotzek,Kohstall,Koschorreck,Zhang,Spethmann,Scazza,Robinson} 
has already led to fundamentally new insights regarding Fermi~\cite{Navon,Punk,Chevy,Cui,Pilati,Massignan1,
Schmidt1,Schmidt2,Ngampruetikorn,Massignan2,Massignan3,Schmidt3,Burovski,Gamayun} 
and very recently Bose polarons~\cite{Palzer,Tempere,Catani,Fukuhara,Scelle,
Schmidtred,Ardila2,Grusdt,Artemis2,Guenther,Mayer}. 
In this latter context the observation of coherent attractive and repulsive quasiparticles~\cite{Jorgensen}, even in the 
strongly interacting regime~\cite{Hu}, refuelled
the scientific interest towards understanding their underlying dynamics. 

Most of the theoretical studies regarding Bose polarons have been mainly focused on 
a mean-field~\cite{Astrakharchik,Cucchietti,Kalas,Bruderer1} 
description and on the Fr{\"o}hlich model~\cite{Sacha,Bruderer,Privitera,Casteels1,Casteels2,Kain}. 
Only very recently theories going beyond the Fr{\"o}hlich paradigm \cite{Li,Ardila1,Shchadilova,Rath,Grusdt1,Lemeshko,Kain1} 
and including higher-order correlations~\cite{Levinsen,Christensen} have been developed thereby allowing for the investigation
of Bose polarons also in the intermediate and strong interaction regime. 
However, current experiments realized both in one~\cite{Catani,Fukuhara,Scelle} and three dimensions~\cite{Jorgensen,Hu}
probed the nonequilibrium dynamics of Bose polarons and necessitated the presence of higher-order correlations
for an adequate description of the observed dynamics.
Thus the interplay of higher-order correlations during the out-of-equilibrium dynamics of bosonic impurities immersed in a 
Bose-Einstein condensate (BEC) is a key ingredient for advancing our understanding of the dynamics of such MB systems.  
On the theoretical side efforts concerning the nonequilibrium dynamics of Bose 
polarons~\cite{Zvonarev,Bonart1,Bonart2,Artemis1,Lampo,Grusdt2} are quite recent and remarkably only 
few of them include quantum fluctuations~\cite{Grusdt2,Mistakidis_bose_pol,Enss_beyond}.

In this Letter, motivated by current experiments~\cite{Jorgensen,Hu,Catani,Cetina0,Cetina} 
we explore the interaction quench dynamics of a spinor impurity coupled to a BEC. 
Focusing on repulsively interacting multi-component bosonic systems in a one-dimensional (1D) harmonic trap, 
we showcase the dynamical dressing of the impurity when all particle correlations are taken into account. 
Three distinct dynamical regions with respect to the interspecies interaction strength are identified and captured by the 
structure factor which is the spin polarization (contrast) of the impurity~\cite{Nishida}. 
These regions are shown to be related to the miscible and immiscible character of the system and are indicative of 
the ohmic character of the bath  \cite{ohmic1,Lampo}. 
Their extent can be manipulated by adjusting the intraspecies repulsion of the BEC alias bath or by changing 
its particle number thereby addressing the few to many-body crossover. 
This tunability is of significant importance since it leads to a longevity of the polaron and thus 
facilitates the control of quasiparticles. 
One of our key results consists of the interpretation of the Bose polaron dynamics in terms of an effective potential. 
The latter is found to be an adequate approximation in the weakly interacting case assuming the Thomas-Fermi approximation 
for the bath and generalizes the results of \cite{Sartori}. 
We demonstrate that deep in the immiscible phase, where entanglement is strong, the Bose polaron ceases to exist due to the 
orthogonality catastrophe~\cite{Anderson,Busch}.
In this strong interaction regime a dissipative motion of the impurity is observed accompanied by 
the population of several lower-lying excited states of the effective potential. 
The latter involves now the single-particle density of the MB bath and provides only a very approximate picture 
of the impurity dynamics since entanglement is significant.   
This mechanism of dissipation in turn leads to a transfer of energy from the impurity to its environment also 
leading to a substantial entanglement in the system. 
\begin{figure}[ht]
\includegraphics[width=1.0\columnwidth]{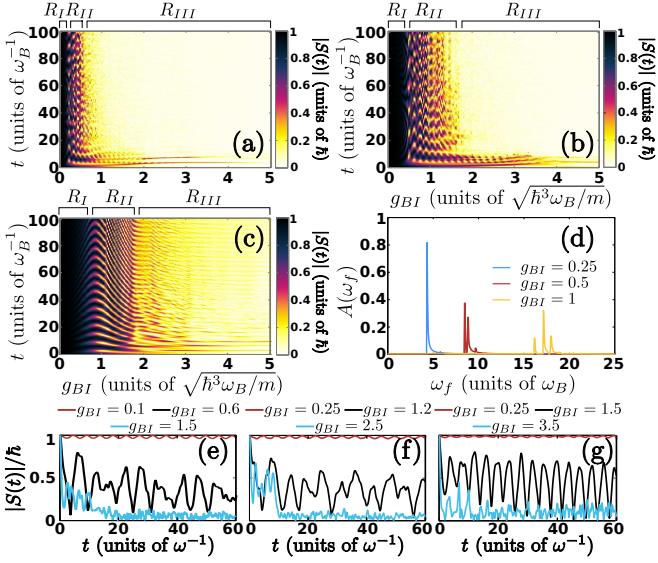}
\caption{(Color online) Evolution of the contrast, $|S(t)|$, upon increasing $g_{BI}$ 
for (a) $g_{BB}=0.2$ and (b) $g_{BB}=0.5$ with $N_{B}=100$ and $N_I=1$. (c) same as (b) but for $N_{B}=10$. 
(d) Excitation spectrum, $A(\omega_f)$, indicating the emergent polaronic peaks for distinct $g_{BI}$ (see legend) and $g_{BB}=0.5$. 
(e), (f), (g) illustrate $|S(t)|$ of (a), (b), (c) for different $g_{BI}$ (see legend).} 
\label{Fig:1}
\end{figure}
\vspace{0.3cm} 

{\it Model.--}
We consider a system consisting of a single impurity of mass $m_{I}$ having an additional spin-$1/2$ degree of freedom. 
The impurity is in the superposition 
$\ket{\Psi_S}=\alpha \ket{\uparrow} + \beta \ket{\downarrow}$, with $\alpha$, $\beta$ denoting the different 
weights used that account for a partial or complete dressing of the single impurity. 
The impurity is immersed in a 1D harmonically confined BEC of $N_B=100$ repulsively interacting atoms of mass $m_{B}$ and trap 
frequency $\omega_B=\omega_I=1.0$. 
The MB Hamiltonian of the system reads 
\begin{eqnarray}
\hat{H} = \hat{H}^{0}_{B}+\sum_a \hat{H}^{0}_a+\hat{H}_{BB}+\hat{H}_{BI}. 
\label{Htot}
\end{eqnarray}
Here, $\hat{H}^{0}_{B}=\int dx~\hat{\Psi}^{\dagger}_{B} (x) \left( -\frac{\hbar^2}{2 m_{B}} \frac{d^2}{dx^2}  
+\frac{1}{2} m_{B} \omega_B^2 x^2 \right) \hat{\Psi}_{B}(x)$ 
is the Hamiltonian describing the motion of the BEC that serves as a bath for the impurity atom. 
$\hat{H}^{0}_a=\int dx~\hat{\Psi}^{\dagger}_a(x) \left(-\frac{\hbar^2}{2 m_I} \frac{d^2}{dx^2}  
+\frac{1}{2} m_I \omega_I^2 x^2 \right) \hat{\Psi}_a(x)$ ($a=\left\{ \uparrow, \downarrow \right \}$) is the 
corresponding Hamiltonian for the impurity atom. 
In both cases $\hat{\Psi}_i (x)$ is the bosonic field-operator of either the majority ($i=B$) or the impurity ($i=a$) atoms. 
We focus on the case of equal masses $m_B=m_I=m$~\cite{Jorgensen}.  
$\hat{H}_{BB}=g_{BB} \int dx~\hat{\Psi}^{\dagger}_{B}(x) \hat{\Psi}^{\dagger}_{B}(x) \hat{\Psi}_{B} (x)\hat{\Psi}_{B}
(x)$ accounts for the contact intraspecies interaction of strength $g_{BB}>0$ in the BEC component. 
$\hat{H}_{BI}=g_{BI}\int dx~\hat{\Psi}^{\dagger}_{B}(x) \hat{\Psi}^{\dagger}_{\uparrow}(x) \hat{\Psi}_{\uparrow} 
(x)\hat{\Psi}_{B}(x)$ denotes the interaction between the bath and the part of the impurity being in the spin-$\uparrow$ 
state, characterized by an effective strength $g_{BI}>0$, while having a noninteracting spin-$\downarrow$ component.
Similar setups have been used in the context of Fermionic impurities mostly 
focusing on the attractive side of interactions~\cite{Chevy1,Combescot,Combescot1,Combescot2,Parishsuper}. 
The multi-component system is initially prepared in its ground state configuration for fixed $g_{BB}$ and $g_{BI}=0$. 
We note that our results remain valid also for the case of weak interspecies interactions.  
Such an initial state preparation is experimentally realizable by means of radiofrequency 
spectroscopy~\cite{Jorgensen,Hu,Cetina,Shchadilova,sgg} and Ramsey interferometry~\cite{Cetina}. 

To derive the nonequilibrium dynamics of the spinor impurity, we use a nonperturbative method, namely
the Multi-Layer Multi-Configuration Time-Dependent Hartree method for atomic mixtures (ML-MCTDHX). 
Our method rests on expanding the MB wavefunction with respect to a variationally optimized time-dependent 
basis which spans the optimal subspace of the Hilbert space at each time instant. 
Its multi-layer ansatz for the total wavefunction allows us to account for all intra- and interspecies 
correlations. 
In our case the latter are found to be more important than the former~\cite{ML,supplemental}. 

Our starting point is the ground state, $\ket{\Psi^0_{BI}}$, obeying the eigenvalue equation 
$\left(\hat{H}-\hat{H}_{BI}\right)\ket{\Psi^0_{BI}}=E_0\ket{\Psi^0_{BI}}$, with $E_0$ denoting the corresponding eigenenergy.
We then abruptly switch on at $t=0$ the interspecies repulsion $g_{BI}$, and let the system evolve dynamically.
The MB wavefunction following the quench reads
\begin{equation}
\ket{\Psi(t)}= \alpha e^{-i\hat{H} t/\hbar} 
\ket{\Psi^0_{BI}} \ket{\uparrow} + \beta e^{-iE_0 t/\hbar} \ket{\Psi^0_{BI}} \ket{\downarrow}.	
\label{Eq:2}
\end{equation} 

{\it Results and Discussion.--} 
To investigate the nonequilibrium dynamics of the spinor impurity 
we first consider the case where the impurity is in an equal superposition
namely $\alpha=\beta=\frac{1}{\sqrt{2}}$,
and determine the time-evolution of the total spin polarization
$|\braket{\hat{\textbf{S}}(t)}|=\sqrt{\braket{\hat{S}_x(t)}^2 + \braket{\hat{S}_y(t)}^2}$.
Here, $\braket{\hat{S}_z(t)}=\braket{\hat{S}_z(t=0)}=0$ since 
$\left[ \hat{S}_z, \hat{H} \right]=0$, while 
$\hat{S}_i=\int dx \sum_{ab} \hat{\Psi}^{\dagger}_a (x) \sigma^i_{ab} \hat{\Psi}_b (x)$
is the spin operator in the $i$th-direction ($i=x,y,z$) and $\sigma^i_{ab}$ are the Pauli matrices. 
This quantity is directly related to the so-called Ramsey response~\cite{Cetina}, namely the structure factor 
which is the time-dependent overlap between the interacting and the noninteracting states 
$|\braket{\Psi^0_{BI}| e^{iE_0 t/\hbar} e^{-i\hat{H} t/\hbar}|\Psi^0_{BI}}|^2=|\braket{\hat{\textbf{S}}(t)}|^2 = 
|S(t)|^2 $~\cite{Nishida}.
$S(t)=|S(t)|~e^{i\phi}$, with $\rm{atan} \phi=\braket{\hat{S}_x} /\braket{\hat{S}_y}$, 
and the Hamiltonian, $\hat{H}$, after the quench, when the impurity is dressed, is given by Eq.~(\ref{Htot}). 
\begin{figure}
\includegraphics[width=0.90\columnwidth]{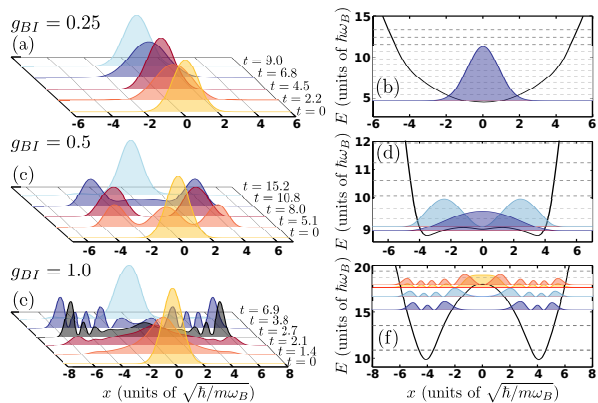}
\caption{(Color online) Selected time instants during evolution of the impurity's one-body density for (a) $g_{BI}=0.25$, (c) 
$g_{BI}=0.5$, and (e) $g_{BI}=1.0$ illustrating its dynamical dressing. 
Effective potential and example densities of the corresponding impurity eigenstates for the aforementioned 
(b) small, (d) intermediate, and (f) large $g_{BI}$ values. 
Notice that the eigenenergies of $V_{eff}$ are slightly shifted with respect to the polaronic energies obtained 
within the MB approach [see also Fig. \ref{Fig:1} (d) and the discussion in the main text]. 
In all cases dashed gray lines correspond to the energy levels of the effective potential.} 
\label{Fig:2}
\end{figure}

Figures~\ref{Fig:1}(a), \ref{Fig:1}(b) and \ref{Fig:1}(c) 
illustrate the evolution of the structure factor $|S(t)|$ (contrast) upon increasing 
the interspecies repulsion $g_{BI}$ for different $g_{BB}$ interactions and also for smaller system sizes.  
In all cases, three distinct dynamical regions can be inferred namely $R_I$, $R_{II}$ and $R_{III}$ which e.g. 
for $g_{BB}=0.5$ correspond to $0\leq g^{R_{I}}_{BI} < 0.5 $, $0.5\leq g^{R_{II}}_{BI} < 1.65 $ 
and $1.65\leq g^{R_{III}}_{BI} < 5.0 $ respectively. 
For short times a descent of $|S(t)|$ is observed~\cite{Cetina,Jalabert}, see 
Figs.~\ref{Fig:1}(e), \ref{Fig:1}(f), and \ref{Fig:1}(g), being sharper for larger $g_{BI}$. 
This descent occurs independently of the value of the intraspecies repulsion $g_{BB}$, compare Figs. \ref{Fig:1}(e), \ref{Fig:1}(f).  
For larger evolution times $|S(t)|$ performs oscillations that become more pronounced upon increasing $g_{BI}$ within 
$R_I$ and exhibit a decaying amplitude in $R_{II}$. 
In contrast entering $R_{III}$ $|S(t)|$ exhibits an exponential decay indicating the orthogonality catastrophe. 
The degree of damping of $|S(t)|$ within $R_{I}$, $R_{II}$ and $R_{III}$ is indicative of a sub-ohmic, ohmic and super-ohmic 
behavior of the bath respectively (see also below). 
Comparing the temporal evolution of $|S(t)|$ for $g_{BB}=0.2$ [Fig.~\ref{Fig:1}(a)] to the one for $g_{BB}=0.5$ [Fig.~\ref{Fig:1}(b)] 
we observe that the extent of the above-mentioned dynamical regions ($R_I$, $R_{II}$, $R_{III}$) can be manipulated by adjusting $g_{BB}$. 
In particular, for larger $g_{BB}$ an enhanced region of finite contrast that enters deeper into the regime of repulsive interspecies 
interactions can be achieved. 
This behavior is supported upon decreasing the number of bath particles to $N_B=10$ [Fig.~\ref{Fig:1} (c)]. 
In the latter few-body scenario coherent oscillations of $|S(t)|$ are observed [see Fig.~\ref{Fig:1} (c) for $0.8<g_{BI}<1.8$] 
leading to a smoothly decreasing contrast as $g_{BI}$ increases \cite{comment_MF}. 
The aforementioned dynamics takes equally place when the initial superposition state of the spinor impurity involves different 
weights for each spinor component. 
This fact can be understood by analytically calculating $|\langle \hat{\textbf{S}}(t)\rangle|_{\alpha,\beta}$ 
when considering different weights $\alpha$ and $\beta$. 
Indeed, it holds $|\langle \hat{\textbf{S}}(t)\rangle|_{\alpha,\beta}=\sqrt{4{\alpha^2}{\beta^2}\abs{S(t)}^2+(\abs{\alpha}^2-\abs{\beta}^2)^2}$, where 
$\abs{S(t)}$ stems from the case $\alpha=\beta=1/\sqrt{2}$. 

As expected the energy spectrum of the impurity is changed upon applying an interaction quench~\cite{Parishsuper}.
To quantify this we determine the Fourier transform of $S(t)$. 
At low impurity densities and weak interspecies interactions $S(t)$ is known to be proportional
to the so-called spectral function of quasiparticles 
$A(\omega_f)=\left(1/\pi\right) {\rm Re}\{\int^{\infty}_0 dt~e^{i \omega_f t}~S(t)\}$~\cite{Cetina,Parishsuper,Nishida,Nozi}.
Figure~\ref{Fig:1}(d) illustrates $A(\omega)$ for different interspecies repulsions ranging from small
($g^{R_{I}}_{BI}=0.25$) to intermediate ($g^{R_{II}}_{BI}=0.5$) and large ($g^{R_{II}}_{BI}=1.0$) interactions respectively.  
The observed peak at small $g_{BI}$ located at $\omega=4.435$ corresponds to the long-time evolution 
of a well-defined repulsive Bose polaron. 
In $R_{II}$ two dominant peaks are imprinted in $A(\omega_f)$ centered at $\omega_1=8.482$ and $\omega_2=8.859$ respectively.
These two peaks correspond to a well-defined quasiparticle dressed, for higher frequencies, by higher-order excitations of the BEC. 
Figures~\ref{Fig:2}(a), \ref{Fig:2}(c) depict the evolution of the impurity's one-body density, 
$\rho_{\uparrow}^{(1)}(x)=\braket{\Psi (t)|\hat{\Psi}^{\dagger}_{\uparrow} (x) \hat{\Psi}_{\uparrow} (x)|\Psi (t)}$, for small 
and intermediate values of $g_{BI}$. 
The observed out-of-equilibrium dynamics of the spinor impurity in both regions $R_I$ and $R_{II}$ can be well-approximated by the dynamics in an effective potential. 
The latter is obtained by considering the bosonic bath as a static potential superimposed to the external 
harmonic trapping of the impurity, namely 
\begin{equation}
V_{eff}=\frac{1}{2} m_B \omega_B^2 x^2+ g_{BI} \rho_B^{(1)}(x), 
\label{Veff}
\end{equation}
where $\rho_B^{(1)}(x)$ is the single-particle density of the BEC at $t=0$. 
It is important to stress that $V_{eff}$ does not take into account the renormalization of the quasiparticle's zero-point energy 
occuring due to its dressing by the bath \cite{Mistakidis_bose_pol}. 
This deficit, however, shifts the eigenspectrum of the impurity in a homogeneous manner and consequently does not affect its dynamics. 
For small $g_{BI}$ and fixed $g_{BB}$ the Thomas-Fermi approximation, i.e. $\rho_B^{(1)}(x)=\frac{1}{g_{BB}}\left( \mu_B-\frac{1}{2} m_B \omega_B^2 x_B^2 \right)$ with 
$\mu_B$ being the chemical potential of the bath, is valid and $V_{eff}=\frac{1}{2} m_B \tilde{\omega}_B^2 x^2+ c$. 
Then $V_{eff}$ is a parabola shifted by $ c\equiv\frac{g_{BI}}{g_{BB}} \mu_B$ 
possessing a modified trapping frequency \cite{Sartori} $\tilde{\omega}_B^2\equiv \left(1- \frac{g_{BI}}{g_{BB}} \right) \omega_B^2< 
\omega_B^2$ [see Fig.~\ref{Fig:2}(b)]. 
In this case the impurity undergoes a breathing motion [Fig.~\ref{Fig:2}(a)]. 
Note that the notion of $V_{eff}$ can be extended to higher dimensions. 
However, relying solely on this approximation we can assess only the frequencies of the emergent dynamical modes  
i.e. the breathing mode, see also \cite{supplemental}. 
Contrary to this an increase of $g_{BI}$ such that $g_{BI}>g_{BB}$ changes this effective potential picture. 
In this case the system enters the immiscible regime and the initial state involves higher-order excitations in 
the effective potential due to the stronger interaction of the impurity with the bosonic bath [Fig.~\ref{Fig:2}(c)]. 
For these intermediate $g_{BI}$ interactions the impurity density develops a two-hump structure being pushed towards the boundaries 
of the bath and favoring a phase-separated state with the BEC which resides around the trap center (see the discussion below). 
It is for these intermediate values, indicating a miscible to an immiscible phase transition, 
that $V_{eff}(x)$ begins to deform into a double-well potential [Fig.~\ref{Fig:2}(d)]. 
The impurity state corresponds then to the ground or the first excited state of this effective potential. 
Further increase of $g_{BI}$ leads to the appearance of three dominant peaks in the impurity's excitation spectrum. 
These peaks are centered at $\omega_3=16.15$, $\omega_4=17.15$, and $\omega_5=17.97$ respectively 
[Fig.~\ref{Fig:1}(d)], and correspond to even higher excited states of the quasiparticle. 
The relevant dynamical evolution of the impurity [Fig.~\ref{Fig:2}(e)] showcases the deformation of its one-body density, 
with these higher excited states occupying the third up to sixth excited state of $V_{eff}$ [Fig.~\ref{Fig:2}(f)].
Entering deeper into the immiscible phase [Fig.~\ref{Fig:1}(b)] results to a fast decay of the contrast at short time
scales. 
Consequently there is no clear polaronic signature in the relevant excitation spectrum but rather a multitude of states are 
occupied in this effective double-well picture. 
This behavior is caused by the dissipative motion of the impurity leading to a partial transfer of its energy to 
the bath as we shall argue below. 
\begin{figure}[ht]
\includegraphics[width=1.0\columnwidth]{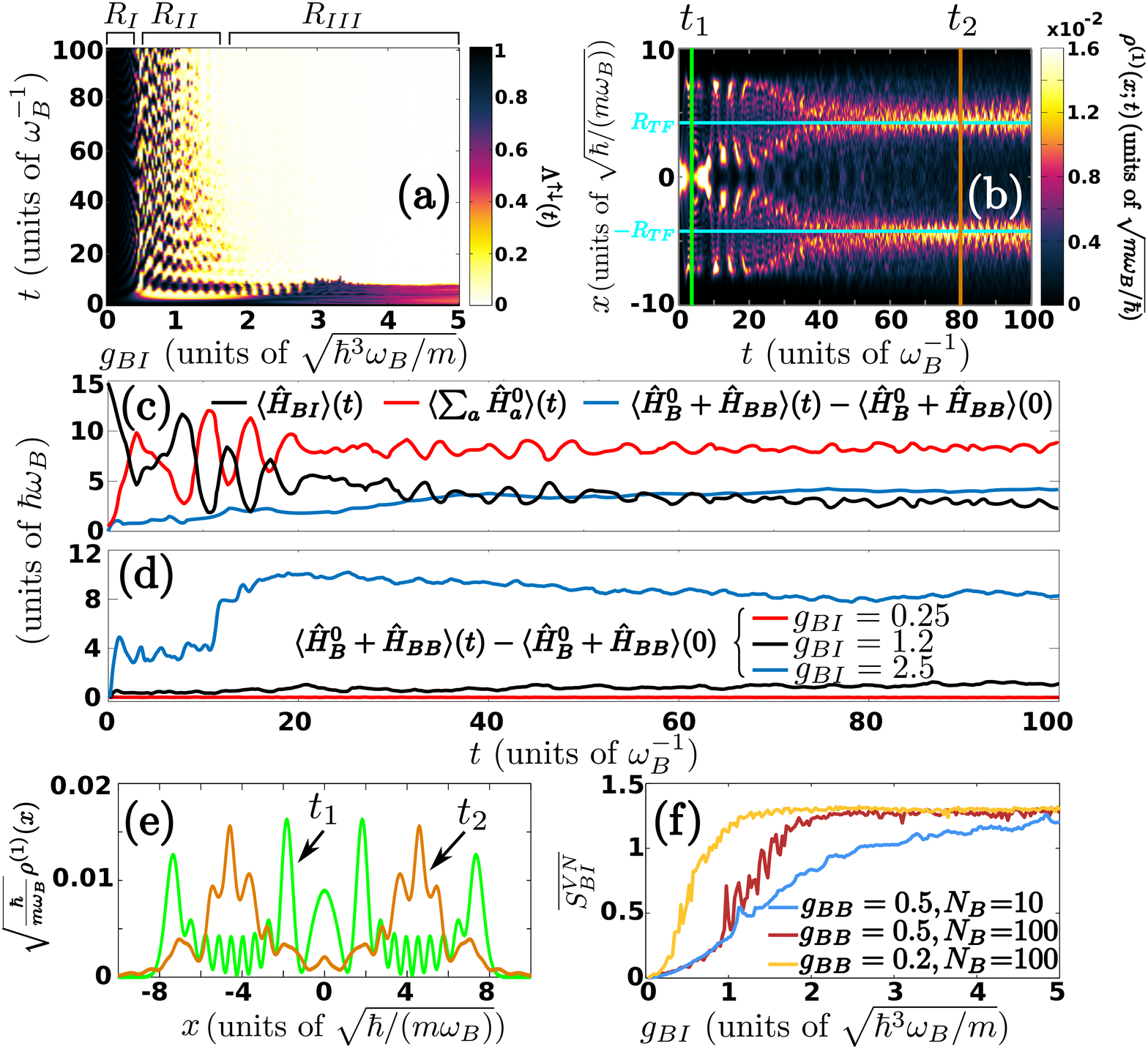}
\caption{(Color online) (a) Evolution of the overlap, $\Lambda^{\uparrow \downarrow}(t)$, between the spin-$\uparrow$ and 
spin-$\downarrow$ states of the impurity atom. 
(b) One-body density evolution of the spin-$\uparrow$ atom. 
Horizontal solid lines indicate the position of the Thomas Fermi radius of the bath. 
(c) Expectation value of the energy (see legend). 
In both (b), (c)  $g_{BI}=1.7$. 
(d) Expectation value of the energy of the bath for different $g_{BI}$ (see legend). 
(e) Density profiles at the time instants marked by the vertical solid lines in (b). 
(f) Time average of the von-Neumann entropy, $\bar{S}_{BI}^{VN}$,  for increasing $g_{BI}$.
In all cases $N_{B}=100$, $N_I=1$ and $g_{BB}=0.5$.} 
\label{Fig:3}
\end{figure}

To deepen our understanding of the dynamics of the spinor impurity we next examine the degree of miscibility 
between the spin components captured by the overlap integral
\begin{eqnarray}
\Lambda^{\uparrow \downarrow}(t)=\frac{\left[\int dx \rho^{(1)}_{\uparrow} (x,t) \rho^{(1)}_{\downarrow} (x,t)\right]^2}
{\int dx \left(\rho^{(1)}_{\uparrow} (x,t)\right)^2 \int dx \left(\rho^{(1)}_{\downarrow} (x,t)\right)^2 }
. 
\label{Lbi}
\end{eqnarray}
Here, e.g. the one-body density of the spin-$\downarrow$ is $\rho^{(1)}_{\downarrow} 
(x,t)=\braket{\Psi(t)|\Psi_{\downarrow}^{\dagger}(t) \Psi_{\downarrow}(t)|\Psi(t)}$. 
$\Lambda^{\uparrow \downarrow}(t)$ takes values within the interval $[0,1]$ with zero (unity) denoting the 
phase immiscible (miscible) spin components. 
Evidently, the three distinct dynamical regions captured by $|S(t)|$ leave their fingerprints 
in $\Lambda^{\uparrow \downarrow}(t)$ [Fig.~\ref{Fig:3}(a)]. 
Note here that $\rho^{(1)}_{\downarrow}(x,t)=\rho^{(1)}_{\uparrow}(x,0)$ and therefore $\Lambda^{\uparrow \downarrow}(t)$ is directly 
related to the contrast [see Fig.~\ref{Fig:1}(b)]. 
Indeed, within $R_I$ the spin components are maximally miscible while within $R_{II}$ they oscillate between miscibility and immiscibility. 
Finally when the orthogonality catastrophe takes place in $R_{III}$ they become immiscible. 
This spin segregation, in $R_{III}$, is manifested in the spatio-temporal evolution of 
$\rho^{(1)}_{\uparrow} (x,t)$ [Fig.~\ref{Fig:3}(b)]~\cite{commentdown}. 
Evidently, $\rho^{(1)}_{\uparrow} (x,t)$ breaks into two density fragments that perform damped oscillations symmetrically placed around 
the edges of the Thomas Fermi radius of the bath. 
These damped oscillations essentially indicate that the spin-$\uparrow$ impurity is initially 
in a highly excited state of $V_{eff}(x)$ [see Fig.~\ref{Fig:3}(e) for $t_1$] while for later times, e.g. $t_2$, 
it populates a superposition of lower excited states. 
We remark here that $\rho^{(1)}_{\uparrow} (x,t)$ depicted in Fig.~\ref{Fig:3}(e) is obtained from the correlated MB calculation while 
the interpretation in terms of $V_{eff}$ provides an approximate picture of the impurity dynamics for these strong interactions. 
The latter behavior implies a transfer of energy from the impurity to the BEC environment [Fig.~\ref{Fig:3}(c)] which is 
beyond the single-particle dynamics provided via $V_{eff}$.  
This energy transfer possesses contributions of different magnitude from each term of the above-mentioned superposition 
leading to different excitations of the BEC and hence it constitutes a manifestation of the entanglement present in the MB system. 
Since the kinetic energy of the impurity increases during evolution also an increase of its non-interacting 
energy, $\langle \sum_{a} {\hat H}^0_a \rangle$, is observed. 
Contrary to this excess of energy, a decrease of the interaction energy $\langle {\hat H}_{BI} \rangle$     
occurs since the impurity is expelled to the edges of the BEC where $\rho^{(1)}_{B}(x)\ll \rho^{(1)}_{B}(0)$. 
Indeed $\langle \hat{H}^0_{B} + \hat{H}_{BB}\rangle$ increases in the course of the dynamics capturing the transfer of energy 
from the impurity to the bath. 
This dissipation mechanism becomes pronounced within $R_{III}$. 
Figure \ref{Fig:3} (d) shows $\langle \hat{H}^0_{B} + \hat{H}_{BB}\rangle$ during evolution for different $g_{BI}$. 
It becomes evident that within $R_I$ the impurity does not dissipate 
energy to the bath since the energy of the latter remains almost constant. 
However, within the region $R_{II}$ the impurity starts to dissipate energy to 
the bath and this dissipation rate becomes maximal within $R_{III}$. 
This observation further supports the sub-ohmic, ohmic and super-ohmic behavior of the bath 
in the different regions. 
Moreover, to directly expose the presence of entanglement with respect to $g_{BI}$ we invoke the 
von-Neumann entropy, $S^{VN}_{BI}(t)=-\sum_i \lambda_i(t) \log \lambda_i(t)$~\cite{Horodeki}. 
Note that $\lambda_i$'s are the eigenvalues of the $N_B$-body density matrix 
$\rho^{(N_B)}_B=-\rm{Tr}_{I} \left[\ket{\Psi(t)} \bra{\Psi(t)} \right]$. 
Indeed, the time average $\bar{S}^{VN}_{BI}$ [Fig.~\ref{Fig:3} (f)] 
shows that the dressed impurity is entangled with the BEC within the regions $R_I$ and $R_{II}$. 
By inspecting $\bar{S}^{VN}_{BI}$ we observe that its slope becomes maximal in $R_{II}$ and therefore 
the same holds for the generation of entanglement, see also \cite{supplemental}. 
Most importantly the system becomes strongly entangled within $R_{III}$, where the polaron ceases to exist,
showcasing a plateau of $\bar{S}^{VN}_{BI} (t)\approx 1.2$ for fixed $g_{BB}=0.5$ and 
for all $g_{BI}\gtrsim 1.65$.

\vspace{0.3cm}
{\it Conclusions.--} 
The correlated quench-induced dynamics of a trapped spinor impurity repulsively interacting with a BEC has been 
investigated. 
Inspecting the evolution of the spin polarization reveals three distinct dynamical regions with respect to the interspecies interaction strength. 
These regions are inherently related to the segregated nature of the multi-component system and can be tuned by changing 
the intraspecies repulsion of the BEC or its particle number thereby addressing the few to many-body crossover.  
Within these three regions the birth, dynamical deformation and death (orthogonality catastrophe) of the Bose polaron 
are unravelled. 
To interpret the impurity dynamics, an effective potential is derived being an adequate approximation 
for weak interspecies repulsions. 
For strong repulsions the system is strongly entangled and the impurity's motion becomes dissipative transferring 
a part of its energy to the bath while being pushed to the edges of the BEC. 
Our results pave the way for manipulating the quasiparticle dynamics. 
An intriguing perspective for future endeavors is to consider more than one impurities where 
induced interactions can play an important role.

\vspace{0.1cm}
\begin{acknowledgements}
S.I.M. and P.S. gratefully acknowledge financial support by the Deutsche Forschungsgemeinschaft 
(DFG) in the framework of the SFB 925 ``Light induced dynamics and control of correlated quantum
systems''. 
G.M.K and P.S. acknowledge the support by the excellence cluster 
`` The Hamburg Center for Ultrafast Imaging: Structure, Dynamics and Control
of Matter at the Atomic Scale'' of the DFG.  

\vspace{0.1cm}
G.C.K, S.I.M. and G.M.K. contributed equally to this work.
\end{acknowledgements}

{}

\newpage~\\
\onecolumngrid
\begin{center}
    \textbf{\large Supplemental Material: Quench Dynamics and Orthogonality Catastrophe of Bose Polarons}
\end{center}
\twocolumngrid
\setcounter{equation}{0}
\setcounter{figure}{0}
\setcounter{table}{0}
\setcounter{page}{1}
\makeatletter
\renewcommand{\theequation}{S\arabic{equation}}
\renewcommand{\thefigure}{S\arabic{figure}}
\renewcommand{\bibnumfmt}[1]{[S#1]}
\renewcommand{\citenumfont}[1]{S#1}

\section{Breathing Dynamics of the Impurity within the Effective Potential Approach} \label{sec:numerics_breathing}

To demonstrate the applicability of our effective potential approach given in Eq. (3) in the main text we investigate the 
breathing dynamics of the single impurity immersed in the bosonic bath. 
In order to capture the breathing dynamics \cite{Siegl,Koutentakis,Ronzheimer,Abraham} in all three dynamical 
regions ($R_I$, $R_{II}$ and $R_{III}$) we consider a quench of the harmonic trapping frequency $\omega_I$. 
Such a process aims at dynamically exciting the corresponding breathing mode of the impurity~\cite{Navon}. 

\begin{figure}[ht]
\includegraphics[width=0.8\columnwidth]{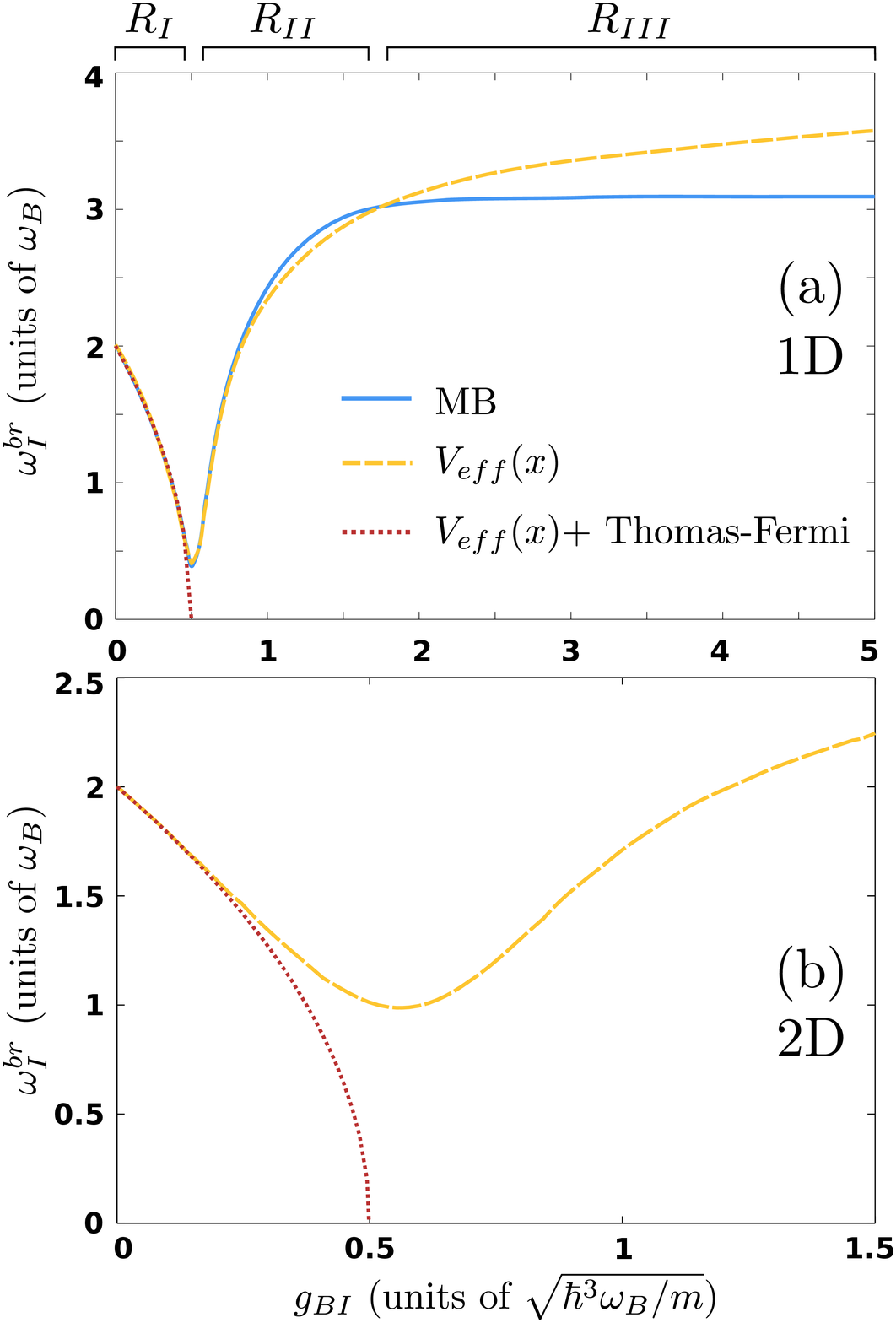}
\caption{(Color online) Breathing frequency, $\omega_I^{br}$, of the Bose polaron as a function of the interspecies 
interaction strength in (a) one-dimension and (b) two-dimensions.  
The breathing motion is induced by quenching the trapping frequency of the impurity from 0.95 to 1.0.  
In all cases $N_{B}=100$, $N_I=1$ and $g_{BB}=0.5$. 
Dotted red line line refers to $\omega_{br}=2 \tilde{\omega}_B$ and dashed yellow line denotes the breathing frequency 
obtained within the effective potential picture (see text).}  
\label{Fig:breathing}
\end{figure}

In particular, we initialize the multi-component system in its ground state for $g_{BB}=0.5$ and a chosen $g_{BI}$ with harmonic oscillator frequencies 
$\omega_B=1$ and $\omega_I=0.95$. 
To trigger the dynamics we suddenly change at $t=0$ the value of $\omega_I$ from 0.95 to 1.0, thus inducing a collective breathing mode. 
Measuring the center-off-mass motion \cite{Siegl,Koutentakis,Ronzheimer} of the impurity we obtain its breathing 
frequency $\omega_I^{br}$ for a specific $g_{BI}$. 
Figure \ref{Fig:breathing} shows $\omega_I^{br}$ for varying $g_{BI}$. 
As it can be seen, $\omega_I^{br}$ decreases within region $R_I$ reaches a critical point around $g_{BI}\approx 0.5$ 
and thereafter it increases within region $R_{II}$ and finally saturates close to $\omega_I^{br}\approx3$ for $g_{BI}>2$. 
We remark here that operating in the mean-field approximation, a similar behavior of $\omega_{I}^{br}$ within the 
miscible phase ($R_I$ region in our case) but for a larger number of impurity particles has been reported in \cite{Sartori}.  

Let us first compare the result of $\omega_I^{br}$ obtained via the many-body (MB) simulations with the effective model 
described in the main text. 
By employing also the Thomas-Fermi approximation the breathing frequency of the impurity reads 
$\omega_{br}=2 \tilde{\omega}_B$ [see the red dotted line in Fig.~\ref{Fig:breathing}] where 
$\tilde{\omega}_B^2=\left( 1-\frac{g_{BI}}{g_{BB}}\right)\omega_B^2$. 
Recall that this effective picture is valid only for weak (miscible) interactions namely when $g_{BB}\ll g_{BI}$.  
We indeed observe that for these interactions the effective potential predicts the correct breathing frequency except for 
the region $g_{BI}\approx g_{BB}=0.5$. 

To extend our analysis to the immiscible regime of interactions where the impurity probes spatial regions beyond the Thomas-Fermi radius, 
we next consider the general form of the effective potential $V_{eff} (x)= 1/2 m \omega_B^2 x^2 + g_{BI} \rho^{(1)}_B(x)$ introduced in 
Eq. (3) of the main text. 
Note that this $V_{eff} (x)$ does not incorporate the Thomas-Fermi approximation. 
Also, $\rho^{(1)}_B(x)$ is the numerically exact ground state density of the bath for $g_{BI}=0$ and $g_{BB}=0.5$. 
To test the accuracy of our effective potential we examine the breathing dynamics of a single particle trapped in $V_{eff} (x)$. 
As before we prepare the impurity in the ground state of $V_{eff} (x)$ with $\omega_I=0.95$ and a specific $g_{BI}$ 
and induce the breathing dynamics by quenching the frequency to 1.0. 
The resulting breathing frequency calculated via the center-of-mass motion is also presented in Fig. \ref{Fig:breathing} (see the yellow dashed line) on top of the 
MB calculation (solid blue line). 
Evidently the single-particle picture obtained within our effective potential provides an adequate approximation 
of the full MB result especially for $0<g_{BI}<1$. 
Deviations between the effective model and the MB calculations are of the order of $8\%$ for $g_{BI}\approx2.5$, while they become 
significant for even larger $g_{BI}$. 
For these strong interactions the entanglement becomes strong rendering the efffective potential an insufficient approach for describing 
the impurity dynamics. 

The notion of the effective potential approximation can be easily extended to higher dimensions. 
Of course, relying exclusively on this approximation it is only possible to access the frequencies of 
the quench-induced dynamical modes, i.e. the breathing mode. 
To showcase whether our predictions of the breathing frequency in one-dimension remain robust in higher dimensions we calculate next 
 the breathing frequency in two-dimensions within the effective potential approximation. 
To induce the breathing dynamics we follow exactly the same procedure as in one-dimension (see the discussion above). 
Assuming an isotropic two-dimensional (2D) external harmonic trap, the effective potential reads 
$V_{eff}(r)=\frac{1}{2} m_B \omega_B^2 r^2+ g_{BI} \rho_B^{(1)}(r)$, 
where $r=\sqrt{x^2+y^2}$ and $\rho_B^{(1)}(r)$ is the single-particle density of the BEC at $t=0$. 
Figure \ref{Fig:breathing} (b) illustrates $\omega_{I}^{br}$ for the 2D 
trapped system upon varying $g_{BI}$.
Additionally, $\rho_B^{(1)}(r)$ is obtained by solving the 2D Gross-Pitaevskii equation. 
For self consistency reasons $\omega_{I}^{br}$ is measured only in regions $R_I$ and $R_{II}$ since already for values of 
$g_{BI}$ that belong to $R_{II}$ the effective potential approximation is expected to fail. 
It is found that $\omega_{br}^I$ exhibits a similar behavior to its one-dimensional counterpart. 
In particular, $\omega_I^{br}$ decreases within region $R_I$, reaches a minimum located around $g_{BI}\approx 0.6$ 
and then it increases deeper in the region $R_{II}$ [see yellow dashed line in Fig.~\ref{Fig:breathing} (b)]. 
Moreover, referring to weak $g_{BI}$ the Thomas-Fermi approximation reads $\rho_B^{(1)}(r)=\frac{1}{g_{BB}}\left( 
\mu_B-\frac{1}{2} m_B \omega_B^2 r_B^2 \right)$.  
Here, $\mu_B$ denotes the chemical potential of the bath. 
Therefore combining the effective potential picture with the Thomas-Fermi approximation 
we deduce that $V_{eff}(r)=\frac{1}{2} m_B \tilde{\omega}_B^2 r^2+ \tilde{c}$, being a parabola shifted 
by $ \tilde{c}\equiv\frac{g_{BI}^{2D}}{g_{BB}} \mu_B$ and 
exhibiting a modified trapping frequency $\tilde{\omega}_B^2\equiv \left(1- \frac{g_{BI}}{g_{BB}} \right) 
\omega_B^2<\omega_B^2$. 
Utilizing this approximation the breathing frequency of the impurity is  
$\omega_{br}=2 \tilde{\omega}_B$ [see the red dotted line in Fig.~\ref{Fig:breathing} (b)]. 
As it can be seen, for $g_{BI}<0.25$ the Thomas-Fermi approximation and the effective potential predict the same 
$\omega_{br}^I$, while for $g_{BI}>0.25$ strong deviations appear. 
Recall that close to $g_{BI}=0.5$ the miscibility/immiscibilty threshold 
is reached and the impurity probes also the spatial 
region at the edge of the BEC density. 
In this region the Thomas-Fermi profile, used herein, is not an adequate approximation for $\rho_B^{(1)}(r)$, 
a result that explains the observed deviations. 
\begin{figure}[ht]
\includegraphics[width=0.9\columnwidth]{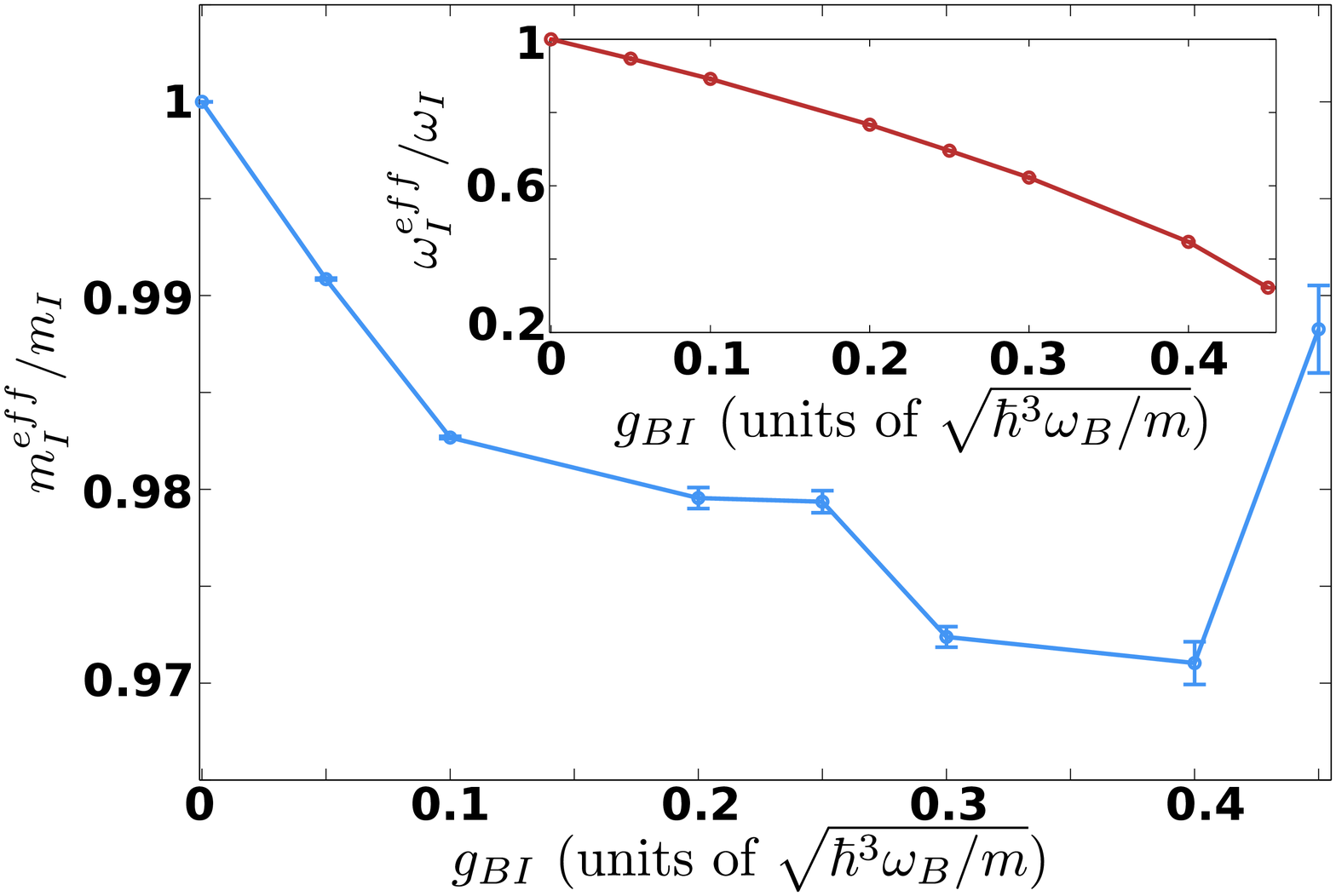}
\caption{(Color online) Effective mass of the polaron for increasing postquench interspecies interaction strength $g_{BI}$. 
The inset shows the effective trapping frequency of the polaron [see also Eq. (\ref{effective})] for varying $g_{BI}$.  
In all cases $N_{B}=100$, $N_I=1$, $g_{BB}=0.5$ and the frequency of the external harmonic confinement is $\omega=1$.  
The system is initialized in its ground state with $g_{BI}=0$ and the dynamics is triggered via an interspecies interaction 
quench to a final value $g_{BI}$.} 
\label{Fig:mass_effective}
\end{figure}

\section{Effective Mass} \label{sec:effective_mass} 

Having at hand the breathing frequency of the impurity atom we next calculate its effective mass $m_{eff}$. 
It has been recently shown \cite{bose_pol} that in the presence of an external harmonic trap the single-particle Hamiltonian 
that governs the impurity dynamics reads 
\begin{equation}
 \hat{H}_I^{eff}=\epsilon_{eff}+\frac{\hat{p}^2}{2m_I^{eff}}+\frac{1}{2}m_I^{eff}(\omega_I^{eff})^2 \hat{x}^2.\label{effective}
\end{equation}
In this expression, $\epsilon_{eff}$ refers to the self-energy of the polaron. 
Also, $\omega_I^{eff}$ denotes the effective trapping of the polaron due to the combined effect of its interaction with the 
bath and the presence of the external harmonic confinement, while $m_I^{eff}$ is the effective mass of the polaron. 
It is also important to note that this effective Hamiltonian description is valid only within the miscible regime 
of interactions since it inherently involves the assumption that the impurity 
is effectively trapped by the bosonic bath. 
For more details regarding the construction of this model we refer the interested reader to \cite{bose_pol}. 
To calculate the effective mass of the polaron within the miscible regime of interactions, referring in our case to $g_{BI}<0.5$, 
we perform the following analysis. 
We first measure the variance (size), $\braket{x^2(t)}$, of the impurity atom for a specific interspecies interaction quench amplitude 
relying on our numerical calculations performed in the main text. 
Independently, by solving Eq. (\ref{effective}) we can show that 
\begin{equation}
\begin{split}
 &\braket{x^2(t)}=\braket{\Psi(t)|\hat{x}^2|\Psi(t)}=\frac{\braket{\Psi(0)|\hat{p}^2|\Psi(0)}}{(m_I^{eff}\omega_I^{eff})^2}\ \\& \times sin^2(\omega_I^{eff}t)+
 \braket{\Psi(t)|\hat{x}^2|\Psi(t)}\cos^2(\omega_I^{eff}t).\label{variance_theory}
\end{split}
 \end{equation}
Assuming an initially non-interacting impurity, i.e. $g_{BI}=0$, we obtain $\braket{\Psi(0)|\hat{p}^2|\Psi(0)}=\frac{\hbar}{2}m_I\omega_I$ and $\braket{\Psi(0)|\hat{x}^2|\Psi(0)}=\frac{\hbar}{2m_I\omega_I}$. 
A similar analytical relation to Eq. (\ref{variance_theory}) can also be obtained for $\braket{\hat{p}^2(t)}$.  
Here the unknown parameters that need to be determined are $\omega_I^{eff}$ and $m_I^{eff}$. 
To estimate these two parameters we perform a fitting of the analytical form of $\braket{x^2(t)}$ given 
by Eq. (\ref{variance_theory}) and $\braket{p^2(t)}$ to the numerically obtained $\braket{x^2(t)}$ and $\braket{p^2(t)}$. 
Figure \ref{Fig:mass_effective} presents $m_I^{eff}$ and $\omega_I^{eff}$, as a result of the above-mentioned fitting, for increasing $g_{BI}$ which 
always lies within the miscible regime of interactions where the polaron is also well defined. 
Recall that for stronger interspecies interactions a dissipative motion of the impurity 
into the bosonic bath takes place  
signalling the onset of the orthogonality catastrophe of the bose polaron. 
As it can be seen in Fig. \ref{Fig:mass_effective}, $m_I^{eff}$ becomes smaller than the bare mass 
of the impurity for increasing interspecies interaction strengths.
A trend that is also followed by $\omega_I^{eff}$ as it can be deduced by inspecting the inset  
depicted in Fig. \ref{Fig:mass_effective}. 
This behavior of $m_I^{eff}$ being in line with the findings of Ref. \cite{bose_pol} 
is attributed to the presence of the external harmonic confinement 
and the interspecies correlations between the bath and the impurity.

\section{The Many-Body Computational Approach: ML-MCTDHX} \label{sec:numerics1}

To simulate the MB quantum dynamics of the composite system discussed in the main text we utilize 
the Multi-Layer Multi-Configuration Time-Dependent Hartree method for Atomic Mixtures
\cite{MLX} (ML-MCTDHX). 
ML-MCTDHX \cite{Lushuai,MLX} is a ab-initio variational method 
for solving the time-dependent MB Schr{\"o}dinger equation of atomic mixtures consisting 
either of bosonic \cite{phassep,darkbright,bose_pol} or fermionic \cite{ferro,fermi_polarons,phase_sep_ferm} species. 
Within this approach the total MB wavefunction is expanded in terms of a time-dependent and variationally 
optimized basis, enabling us to capture the important correlation effects by using a computationally 
feasible basis size. 
In this way the system relevant subspace of the Hilbert space is spanned in an efficient manner at each time 
instant using a reduced number of basis states when compared to expansions relying on a time-independent basis. 
Most importantly, its multi-layer structure allows for tailoring the MB wavefunction ansatz to account for 
both intra- and interspecies correlations when simulating the dynamics of composite systems. 
Due to the above ML-MCTDHX constitutes a versatile tool for simulating the dynamics 
of multispecies systems. 

Here, we employ ML-MCTDHX in order to study the quench-induced correlated dynamics of a particle imbalanced 
bosonic mixture. 
The mixture consists of a majority species being referred to as bath (B) in the following and a 
minority species which we shall call impurity (I) below. 
Most importantly, the minority atoms possess an additional spin-$1/2$ degree of freedom. 
To account for inter- and intraspecies correlations, the MB wavefunction ($|\Psi(t)\rangle$) is firstly expressed 
as a linear combination of $D$ time-dependent species wavefunctions ($|\Psi^{\sigma}_i(t)\rangle$) for each of the 
$\sigma=B,I$ species  
\begin{equation} 
    |\Psi(t)\rangle=\sum_{i,j=1}^D A_{ij}(t) |\Psi^{\rm B}_i(t)\rangle|\Psi^{\rm I}_j(t)\rangle. 
    \label{eq:tot_wfn}
\end{equation}
Here $A_{ij}(t)$ refer to the corresponding time-dependent expansion coefficients. 
We remark that Eq. (\ref{eq:tot_wfn}) is connected to the truncated Schmidt decomposition of rank $D$ \cite{Horodeckix4,darkbright,phassep} 
via a unitary transformation, with the eigenvalues of $A_{ij}(t)$ being the well-known 
Schmidt weights, $\sqrt{\lambda_i(t)}$. 
Following this unitary transformation, $U$, we obtain $A_{ij}=U_{ik}^{-1}\sqrt{\lambda_k}U_{kj}$. 
Then the MB wavefunction is expressed in terms of different interspecies modes of entanglement taking the form 
$|\Psi(t)\rangle=\sum_{k=1}^D \sqrt{\lambda_k(t)} |\tilde{\Psi}^{\rm B}_k(t)\rangle|\tilde{\Psi}^{\rm I}_k(t)\rangle$ with 
$\sqrt{\lambda_k(t)} |\tilde{\Psi}^{\rm B}_k(t)\rangle|\tilde{\Psi}^{\rm I}_k(t)\rangle$ being referred to as the $k$-th mode of entanglement. 
\begin{figure}[ht]
\includegraphics[width=0.9\columnwidth]{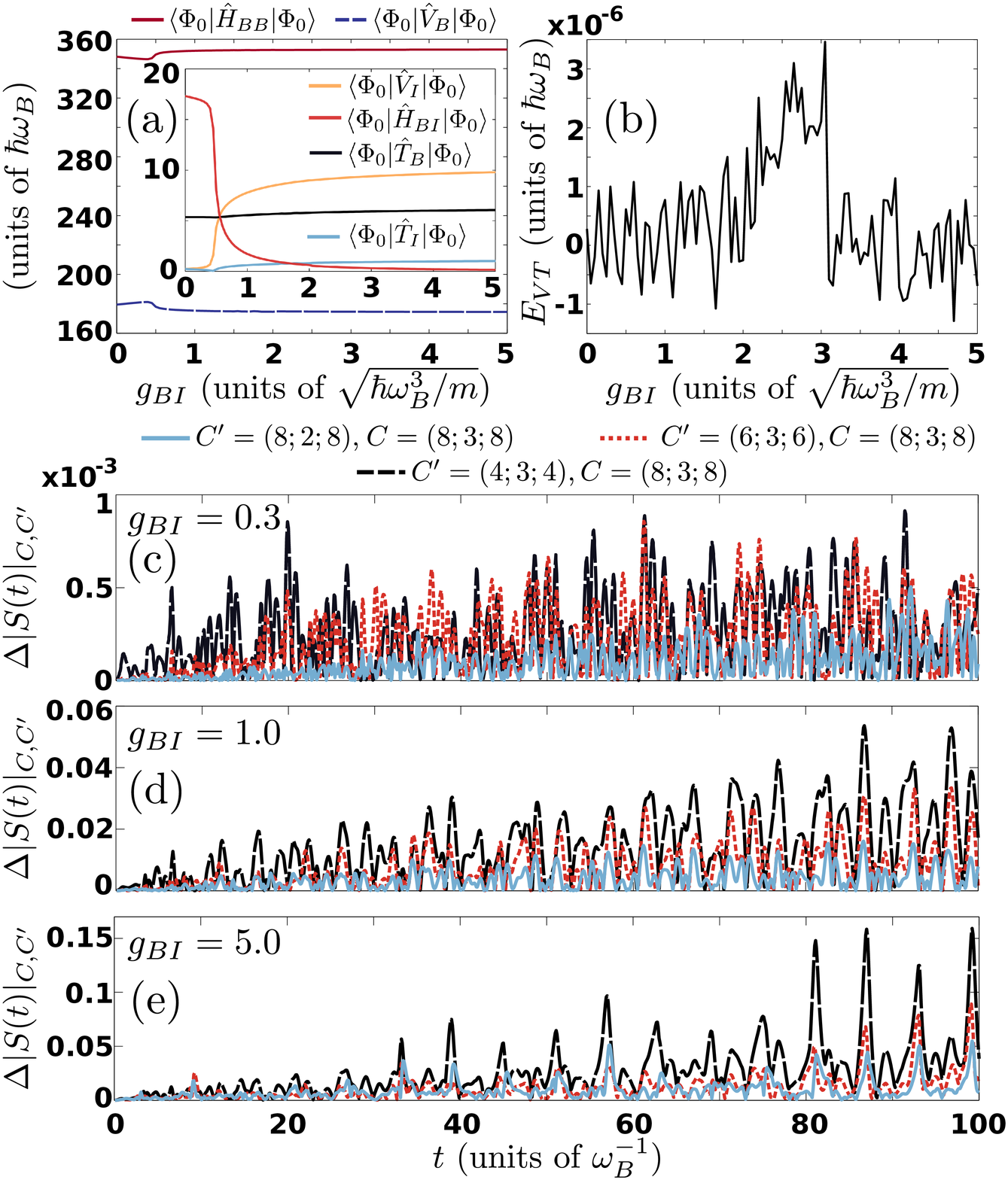}
\caption{(Color online) (a) Different energy contributions (see legend) of the ground state of the multicomponent system for 
varying interspecies interaction strength $g_{BI}$. 
(b) Outcome of the Virial theorem [see also Eq. (\ref{virial})] for increasing $g_{BI}$, obtained via the energy contributions shown in (a). 
We observe that $E_{VT}=0$ for every $g_{BI}$ verifying the Virial theorem for the ground states of our system setups. 
Evolution of the spin polarization deviations $\Delta\abs{S(t)}_{C,C'}$ between the $C=(8;3;8)$ and other orbital configurations $C'=(D;d^A;d^B)$ (see legend) for 
(c) $g_{BI}=0.3$, (d) $g_{BI}=1$ and (e) $g_{BI}=5$. 
In all cases $N_{B}=100$, $N_I=1$ and $g_{BB}=0.5$.} 
\label{Fig:convergence_figure}
\end{figure}

Next each species wavefunction is expanded on the time-dependent number-state 
basis, $|\vec{n} (t) \rangle^{\sigma}$, as   
\begin{equation}
    | \Psi_i^{\sigma} (t) \rangle =\sum_{\vec{n}} B^{\sigma}_{i;\vec{n}}(t) | \vec{n} (t) \rangle^{\sigma}, 
    \label{eq:mb_wfn}
\end{equation}
where $B^{\sigma}_{i;\vec{n}}(t)$ refer to the time-dependent coefficients.  
Each $|\vec{n} (t) \rangle^{\sigma}$ corresponds to a permanent of the $d^{\sigma}$ time-dependent variationally 
optimized single-particle functions (SPFs) denoted by $\left|\phi_l^{\sigma} (t) \right\rangle$, $l=1,2,\dots,d^{\sigma}$ with occupation
numbers $\vec{n}=(n_1,\dots,n_{d^{\sigma}})$. 
The SPFs are subsequently expanded within a time-independent primitive basis. 
For the majority species this primitive basis $\lbrace \left| k \right\rangle \rbrace$ consists of 
an $\mathcal{M}$ dimensional discrete variable representation (DVR). 
Regarding the impurity the primitive basis $\lbrace \left| k,s \right\rangle \rbrace$ corresponds to 
the tensor product of the DVR basis for the spatial degrees of freedom and the two-dimensional 
spin-1/2 basis $\{\ket{\uparrow}, \ket{\downarrow}\}$. 
In this way 
\begin{equation}
    | \phi^{\rm I}_j (t) \rangle= \sum_{k=1}^{\mathcal{M}} \sum_{\alpha=\{ \uparrow, \downarrow \}} C^{{\rm
    I}}_{jk \alpha}(t) \ket{k} \ket{\alpha}, 
    \label{eq:spfs}
\end{equation}
where $C^{{\rm I}}_{j k \alpha}(t)$ are time-dependent expansion coefficients. 
We remark that each SPF for the impurity atoms is a spinor wavefunction $| \phi_j^{\rm I} (t) \rangle= \int {\rm d}x~ [
    \chi_j^\uparrow(x) \hat{\Psi}^\dagger_\uparrow(x) + \chi_j^\downarrow(x)
\hat{\Psi}^\dagger_\downarrow(x) ] | 0\rangle$ (for more details see also \cite{ferro}). 
To obtain the time-evolution of the ($N_B+N_I$)-body wavefunction $\left|\Psi(t) \right\rangle$ under the influence of the 
Hamiltonian $\hat{H}$ [see Eq. (1) in the main text] we determine the equations of motion \cite{MLX} for the 
coefficients ${{A_{\vec n }}(t)}$, $B^{\sigma}_{i;\vec{n}}(t)$ and $C^{{\rm I}}_{jk \alpha}(t)$ by employing 
e.g. the Dirac-Frenkel \cite{Frenkel,Dirac} variational principle for the generalized ansatz [see Eqs.~(\ref{eq:tot_wfn}), (\ref{eq:mb_wfn})]. 
The latter refer to a set of $D^2$ ordinary (linear) differential equations of motion for the ${{A_{\vec n }}(t)}$  
coefficients, coupled to a set of $D(\frac{(N_B+d^B-1)!}{N_B!(d^B-1)!}+\frac{(N_I+d^I-1)!}{N_I!(d^I-1)!})$ 
non-linear integrodifferential equations for the species functions, and $d^B+d^I$ nonlinear integrodifferential equations 
for the SPFs. 
Finally, let us mention that ML-MCTDHX is able to operate within different approximation orders, for instance it reduces to the set of coupled mean-field (MF)
Gross-Pitaevskii equations \cite{darkbook,stringari} when $D=d^B=d^I=1$. 
Moreover it is capable of operating within the species mean-field (SMF) approximation \cite{MLX,Lushuai,darkbright,phassep,bose_pol} 
in which the entanglement between the species is neglected but intraspecies correlations are taken into account. 
In particular, in this latter case the $N_{\sigma}$-body state of each species is described by only one species 
function ($|\Psi^{\rm B}_i(t)\rangle=|\Psi^{\rm I}_i(t)\rangle=0$ for $i\neq1$) building upon distinct single-particle 
functions $\left|\phi_{j}^{B} (t) \right\rangle$ and $\left|\phi_{k}^{I} (t) \right\rangle$ with $j=1,2,\dots,d^{B}$ 
and $k=1,2,\dots,d^{I}$ respectively. 
Accordingly the total wavefunction of the system becomes $\ket{\Psi(t)}=\ket{\Psi_1^B(t)}\otimes \ket{\Psi_1^I(t)}$.  

Within our implementation a sine discrete variable representation (sine-DVR) has been used as the primitive basis for the 
spatial part of the SPFs including $\mathcal{M}=450$ grid points. 
The sine-DVR intrinsically introduces hard-wall boundaries at both edges of the numerical grid being in our case $x_{\pm}=\pm40$.
We have ensured that the location of these boundary conditions does not affect our results since we do not observe appreciable 
densities to occur beyond $x_{\pm}=\pm25$. 
Another way to confirm the absence of edge effects
in the simulations that were carried out and presented in the main text, 
is to estimate the time at which excitations travel towards the boundaries as detected 
by the local speed of sound of the bath. 
Indeed, all possible emergent system correlations will definitely travel with a speed smaller than the local speed of sound.
The relevant time window for travelling a distance from the trap center to the spatial point $x_b>0$ is $T=\int_0^{x_b} \frac{dx}{c(x)}$, where the local 
speed of sound reads $c(x)=\sqrt{\frac{g_{BB}\rho_B^{(1)}(x)}{m}}$. 
This time increases dramatically when $x_b$ lies beyond the Thomas-Fermi radius, which is in our case $R_{TF}\approx 4.2$. 
For instance, when $x_b=6$ we obtain $T\approx106$ while for $x_b\equiv x_{+}=+40$ we find $T\approx2\times10^{16}$. 
Therefore we can again deduce that within the considered simulation time $T=100$, in the main text, edge effects do not play any role. 
To obtain the eigenstates of the MB system we rely on the so-called improved relaxation method~\cite{MLX,Lushuai} within ML-MCTDHX. 
To track the dynamics of the composite bosonic system we propagate in time the wavefunction [Eq.~(\ref{eq:tot_wfn})] by employing 
the appropriate Hamiltonian within the ML-MCTDHX equations of motion.  

To conclude upon the reliability of our results, we increase the number of species functions $D$, 
SPFs $d^B$ and $d^I$, and grid points $\mathcal{M}$, thus observing a systematic convergence of all 
the observables of interest, e.g. $S(t)$ and $\Lambda^{BI}(t)$. 
The Hilbert space truncation, i.e. the order of the used approximation, 
is designated by the considered orbital configuration space $C=(D;d^B;d^I)$. 
Convergence here means that for an increasing orbital configuration $C$ the observables become almost insensitive within 
a given relative accuracy. 
We remark that all MB calculations presented in the main text refer to the 
configuration $C=(8;3;8)$. 

In order to test from first principles the convergence of our results regarding the ground state 
of our system when varying $g_{BI}$ we resort to the quantum Virial theorem. 
Referring to the ground state of our system, $\ket{\Psi(0)}$, the Virial theorem reads
\begin{equation}
\begin{split}
E_{VT}\equiv 2(\braket{\Psi(0)|\hat{T}_B|\Psi(0)}+\braket{\Psi(0)|\hat{T}_I|\Psi(0)})\\-2(\braket{\Psi(0)|\hat{V}_B|\Psi(0)}+\braket{\Psi(0)|\hat{V}_I|\Psi(0)})\\+
 (\braket{\Psi(0)|\hat{H}_{BB}|\Psi(0)}+\braket{\Psi(0)|\hat{H}_{BI}|\Psi(0)})=0.\label{virial}
\end{split}
 \end{equation}
Here, the kinetic energy operators of the bath and the impurity are denoted by $\hat{T}_B=-\int dx \hat{\Psi}_B^{\dagger}(x)\frac{\hbar^2}{2 m_{B}} \frac{d^2}{dx^2}\hat{\Psi}_B(x)$ 
and $\hat{T}_I=-\int dx \hat{\Psi}_I^{\dagger}(x) \frac{\hbar^2}{2 m_{I}} \frac{d^2}{dx^2} \hat{\Psi}_I(x)$ respectively.  
Also, the corresponding potential energy operator for the bath is $\hat{V}_B=\int dx \hat{\Psi}_B^{\dagger}(x)\frac{1}{2} m_{B} \omega_B^2 x^2\hat{\Psi}_B(x)$ 
and for the impurity reads $\hat{V}_I=\int dx \hat{\Psi}_I^{\dagger}(x)\frac{1}{2} m_{I} \omega_I^2 x^2\hat{\Psi}_I(x)$. 
Furthermore, the operator of the intraspecies interaction energy of the bath is 
$\hat{H}_{BB}=g_{BB} \int dx~\hat{\Psi}^{\dagger}_{B}(x) \hat{\Psi}^{\dagger}_{B}(x) \hat{\Psi}_{B} (x)\hat{\Psi}_{B}
(x)$, while the operator of the interspecies interaction energy is  
$\hat{H}_{BI}=g_{BI}\int dx~\hat{\Psi}^{\dagger}_{B}(x) \hat{\Psi}^{\dagger}_{\uparrow}(x) \hat{\Psi}_{\uparrow} 
(x)\hat{\Psi}_{B}(x)$. 
In all cases, $\hat{\Psi}_i (x)$ refers to the bosonic field operator of either the bath ($i=B$) or the impurity ($i=\uparrow,\downarrow$) atoms. 
The different energy contributions, given by Eq. (\ref{virial}), 
are illustrated in Fig. \ref{Fig:convergence_figure} (a) for increasing $g_{BI}$. 
Calculating the quantum Virial theorem by taking into account all the aforementioned contributions 
[see Eq. (\ref{virial})] confirms that it is indeed fulfilled, since $E_{VT}\sim 10^{-6}$, 
for all the ground states of the system independently of $g_{BI}$ [see Fig. \ref{Fig:convergence_figure} (b)]. 
The latter assures the convergence of all the ground state configurations considered herein. 

As a next step, we also briefly discuss the convergence of our results 
in terms of a different number of species and single-particle functions.  
For this investigation we monitor during the nonequilibrium dynamics e.g. the spin polarization, $\abs{S(t)}_C$,  
and we calculate its absolute deviation between the $C=\left(8;3;8\right)$ 
and other numerical configurations $C'=\left(D;d^A;d^B\right)$  
\begin{equation}
\Delta\abs{S(t)}_{C,C'} =\frac{|\abs{S(t)}_{C} -\abs{S(t)}_{C'}|}{\abs{S(t)}_{C}}. \label{converg_test} 
\end{equation} 

Figure \ref{Fig:convergence_figure} presents $\Delta\abs{S(t)}_{C,C'}$ for the multicomponent system under consideration with $N_B=100$ and $N_I=1$ 
following an interspecies interaction quench from $g_{BI}=0$ to different finite values of $g_{BI}$. 
Closely inspecting Fig. \ref{Fig:convergence_figure}, it becomes apparent that a systematic convergence of $\Delta\abs{S(t)}_{C,C'}$ is achieved for all 
interspecies interactions used in the main text. 
For instance, comparing $\Delta\abs{S(t)}_{C,C'}$ at $g_{BI}=0.3$ between the $C=(8;3;8)$ and $C'=(6;3;6)$ [$C'=(8;2;8)$] approximations we deduce that 
the corresponding relative difference lies below $0.08\%$ [$0.04\%$] throughout the evolution, see Fig. \ref{Fig:convergence_figure} (a). 
However, for increasing $g_{BI}$ the relative errors become larger, see Figs. \ref{Fig:convergence_figure} (b), (c). 
E.g. for $g_{BI}=5$ which is the strongest interaction considered in the main text $\Delta\abs{S(t)}_{C,C'}$ between the 
configurations $C=(8;3;8)$ and $C'=(6;3;6)$ [$C'=(8;2;8)$] exhibits a deviation which reaches a maximum value of the order of 
$7\%$ [$5\%$] at large propagation times. 
Finally, we remark that a similar analysis has been performed for all other interspecies interaction strengths discussed within the main text and found 
to be adequately converged (results not shown here for brevity).

\section{Fragmentation} \label{sec:frgmentation} 

The spectral representation of the reduced $\sigma$ species one-body density matrix \cite{Naraschewski} reads 
\begin{equation}
 \rho^{(1)}_{\sigma} (x,x';t) = N_{\sigma} \sum\limits_{\alpha=1}^{d^{\sigma}}
{{n_{\alpha}^{\sigma}}(t){\varphi _{\alpha}^{\sigma}}(x,t)} \varphi _{\alpha}^{* \sigma}(x',t), \label{eq:orbitals}
\end{equation}
where ${\varphi _\alpha^{\sigma}}(x,t)$ are the so-called natural orbitals of the $\sigma=B,I$ species and $d^{\sigma}$  
corresponds to the considered number of orbitals for the $\sigma$ species. 
The corresponding population eigenvalues $n_{\alpha}^{\sigma}(t) \in [0,1]$ (natural populations) characterize the degree of 
intraspecies correlations or fragmentation of the system \cite{darkbright,phassep,Mueller,Penrose}. 
Here we consider the natural orbitals to be normalized to unity i.e. $\int d x \abs{\varphi^{\sigma}_{\alpha}(x)}^2=1$. 
Indeed, for only one macroscopically occupied orbital the system is said to be 
condensed, otherwise it is termed fragmented. 
It can be shown that for $n_1^{\sigma}(t)=1$, $n_{i\neq1}^{\sigma}(t)=0$ 
the first natural orbital $\sqrt{N_{\sigma}}\varphi^{\sigma}_1(x^{\sigma};t)$ reduces to the MF wavefunction 
$\varphi^{\sigma}(x^{\sigma};t)$. 
Therefore, $1-n_1^{\sigma}(t)$ offers a measure of the degree of the $\sigma$ species 
fragmentation \cite{Mueller,Penrose}. 

Finally let us remark that by employing the Schmidt decomposition of Eq. (\ref{eq:tot_wfn}) and the fact that $N_I=1$ the 
one-body density matrix of the impurity reads 
\begin{equation}
 \rho^{(1)}_{I}(x,x';t)=\sum_{k=1}^{D} \lambda_k(t) \tilde{\Psi}_k^I(x,t) \tilde{\Psi}_k^{I*}(x',t),\label{eq:orb_species} 
\end{equation}
where $\tilde{\Psi}_k^{I}(x,t)=\braket{x|\tilde{\Psi}_k^I}$. 
Thus comparing Eq. (\ref{eq:orbitals}) with Eq. (\ref{eq:orb_species}) we can easily deduce that $n_k^I(t)=\lambda_k(t)$ for every $k$.

\section{Manifestation of Entanglement in the Degree of Miscibility} \label{sec:entanglement}

Let us now elaborate on the relation between the degree of miscibility and the 
entanglement occurring among the bath and the spinor impurity. 
As already discussed in the main text in order to expose the degree of phase separation, 
namely the degree of miscibility or immiscibility, between the bath and the spinor impurity 
we invoke the overlap integral function~\cite{jain,Bandyopadhyay} which reads 
\begin{eqnarray}
\Lambda^{BI}(t)=\frac{\left[\int dx ~\rho ^{(1)}_{B}(x;t)  \rho ^{(1)}_{I}(x;t) 
\right]^2}{\int dx \left(\rho ^{(1)}_{B}(x;t) \right)^2 \times 
\int dx \left(\rho ^{(1)}_{I} (x;t) \right)^2}. 
\label{overlap_int}
\end{eqnarray}
In this expression, $\rho ^{(1)}_{\sigma}(x;t)$ denotes the one-body density of the $\sigma=B,I$ species. 
This function is normalized to unity taking values between $\Lambda=0$ and $\Lambda=1$ that signify 
complete or zero spatial overlap respectively on the single-particle level. 
Moreover according to our MB wavefunction expansion [see Eq.~(\ref{eq:tot_wfn})], the one-body 
density of the $\sigma$ species, $\rho^{(1)}_{\sigma}(x;t)$, can be expressed with respect 
to different entanglement modes \cite{darkbright,phassep} as    
\begin{equation}
\begin{split}
\rho^{(1)}_{\sigma}(x;t)=
\sum_{k=1}^D \lambda_k(t)~ \rho^{(1),\sigma}_k(x;t).
\end{split} \label{Eq:density_modes}
\end{equation} 
Here $\rho^{(1),\sigma}_k(x;t)=\braket{\tilde{\Psi}_k^{\sigma}|\hat{\Psi}_{\sigma}^{\dagger}(x)\hat{\Psi}_{\sigma}(x)|\tilde{\Psi}_k^{\sigma}}$ 
is the one-body density of the $k$-th species function. 
$\hat{\Psi}^{\dagger}_{\sigma}(x)$ [$\hat{\Psi}_{\sigma}(x)$] denotes the bosonic field operator that creates (annihilates) a 
$\sigma$ species boson at position $x$. 
Moreover, $\lambda_k(t)$ refer to the corresponding Schmidt coefficients of the truncated Schmidt decomposition [see also Eq.~(\ref{eq:tot_wfn})]. 
We remark that the $\lambda_k$'s in decreasing order are known as natural species populations of the $k$-th species wavefunction $|\Psi^{\rm \sigma}_k(t)\rangle$ 
of the $\sigma$-species. 
In turn, they represent a measure of the entanglement or interspecies correlations between the bath and the impurity. 
In particular, the system is said to be entangled \cite{roncaglia,darkbright,phassep} when at least two different $\lambda_k$'s are nonzero. 
Recall that in this latter case the total MB state [Eq. (\ref{eq:tot_wfn})] cannot be expressed as a direct product of two species states. 

To proceed we define the general overlap integral between the one-body densities of different species wavefunctions corresponding to distinct modes of 
entanglement ($i\neq j$) and species ($\sigma \neq \sigma'$) as 
\begin{equation}
\begin{split}
K_{ij}^{\sigma \sigma'}(t)=\int dx ~\rho^{(1),\sigma}_i(x;t) \rho^{(1),\sigma'}_j(x;t). 
\end{split} \label{Eq:spatial_overlap_entanglement}
\end{equation} 
Note that within the MF as well as the SMF approximations $\rho^{(1),\sigma}_i(x;t)=0$ for $i>1$ holds by definition, since 
entanglement is ignored. 
In these non-entangled limits only $K_{11}^{\sigma \sigma'}$ acquires a non-zero value and hence it is relevant. 
Turning to a full MB description where several species wavefunctions are considered, and therefore entanglement is present, $K_{ij}^{\sigma \sigma'}\neq 0$ 
as long as there is a finite spatial overlap between the different species wavefunctions of the same ($\sigma=\sigma'$) or distinct ($\sigma \neq \sigma'$) species. 
Inserting now Eq. (\ref{Eq:density_modes}) into Eq. (\ref{overlap_int}) and using Eq. (\ref{Eq:spatial_overlap_entanglement}) we can re-express the overlap 
integral in terms of the Schmidt coefficients $\lambda_i$'s and $K_{ij}^{\sigma \sigma'}$. 
Indeed we obtain 
\begin{widetext}
\begin{eqnarray}
\Lambda^{BI}(t)=\frac{[\sum_{i=1}^{D}\lambda_i^2 K_{ii}^{BI}+\sum_{i\neq j} \lambda_i \lambda_j K_{ij}^{BI}]^2 }
{[\sum_{i=1}^{D}\lambda_i^2 K_{ii}^{BB}+2\sum_{i<j} \lambda_i \lambda_j K_{ij}^{BI}] 
[\sum_{i=1}^{D}\lambda_i^2 K_{ii}^{II}+2\sum_{i<j} \lambda_i \lambda_j K_{ij}^{IB}]}.
\label{overlap_general}
\end{eqnarray}
\end{widetext}
Of course in this most general case where a full MB description is considered and entanglement is strong, namely more than one species wavefunctions 
are significantly occupied, the relation of $\Lambda^{BI}(t)$ with the Schmidt coefficients $\lambda_i$ where $i=1,2,\dots,D$ is complicated. 
To get a better insight of the aforementioned relation let us consider the following limiting cases. 
Within the MF and SMF approximations where entanglement between the bath and the impurity is absent (see also our discussion above) 
it can be easily shown that $\Lambda^{BI}(t)$ becomes 
\begin{eqnarray}
\Lambda^{BI}(t)=\Lambda_{0}(t)\equiv \frac{[K_{11}^{BI}]^2}{K_{11}^{BB}K_{11}^{II}},  
\label{overlap_MF}
\end{eqnarray} 
as there is only a single (and hence dominant) mode in the corresponding Schmidt decomposition. 
Moreover when considering the weakly entangled case where $\lambda_1\approx1$ and $\lambda_j\ll1$ with $j=2,3,\dots,D$ the overlap integral can 
be written with respect to the higher-order Schmidt coefficients as 
\begin{widetext}
\begin{eqnarray}
\Lambda^{BI}(t)=\Lambda_0(t)\left[ 1+2\sum_{j>1} \frac{\lambda_j(t)}{\lambda_1(t)}\left(\frac{K_{1j}^{BI}}{K_{11}^{BB}}+\frac{K_{j1}^{BI}}{K_{11}^{BB}}-\frac{K_{1j}^{II}}{K_{11}^{II}}
-\frac{K_{1j}^{BB}}{K_{11}^{BB}}\right)\right] +\mathcal{O}\left((\frac{\lambda_j}{\lambda_1})^2\right).
\label{overlap_weak_entanglement}
\end{eqnarray}
\end{widetext} 
We remark here that this weakly entangled case is actually realistic in a MB treatment only within the miscible phase since 
immiscible species are strongly entangled. 
It becomes evident that in this weakly entangled case the major contribution of $\Lambda^{BI}(t)$ stems from the dominant mode described by $\lambda_1$ 
and being encrypted in $\Lambda_0(t)$. 
However, there are small additional contributions to $\Lambda^{BI}(t)$ being of the order of $\lambda_j/\lambda_1$. 
These latter contributions originate from the overlap between the one-body densities of the first [$\rho^{(1),\sigma}_1(x;t)$] 
with the higher-order [$\rho^{(1),\sigma}_j(x;t)$, $j>1$] modes of entanglement. 
It is important to stress that the species wavefunction of the first mode, $\Psi_1^{\sigma}(t)$, and therefore its density $\rho^{(1),\sigma}_1(x;t)$ is greatly 
altered in the full MB case when compared to the MF and SMF cases where entanglement is neglected. 
Concluding, $\Lambda^{BI}(t)$ captures the manifestation of entanglement even in such a weakly entangled scenario, while more traditional measures such as the 
Von-Neumann entropy $S^{VN}_{BI}=-\sum_i \lambda_i(t) \log \lambda_i(t)$ are not sensitive to this change. 
Indeed $S^{VN}_{BI}$ depends on the number and the weights ($\lambda_i$'s) of the entanglement modes. 
Therefore $\Lambda^{BI}(t)$ greatly supplements $S^{VN}_{BI}$ regarding the identification of entanglement induced effects.

\section{Generation of Entanglement and Fragmentation versus the Degree of Miscibility} \label{sec:entanglement_fragmentation}

\begin{figure}[ht]
\includegraphics[width=1.0\columnwidth]{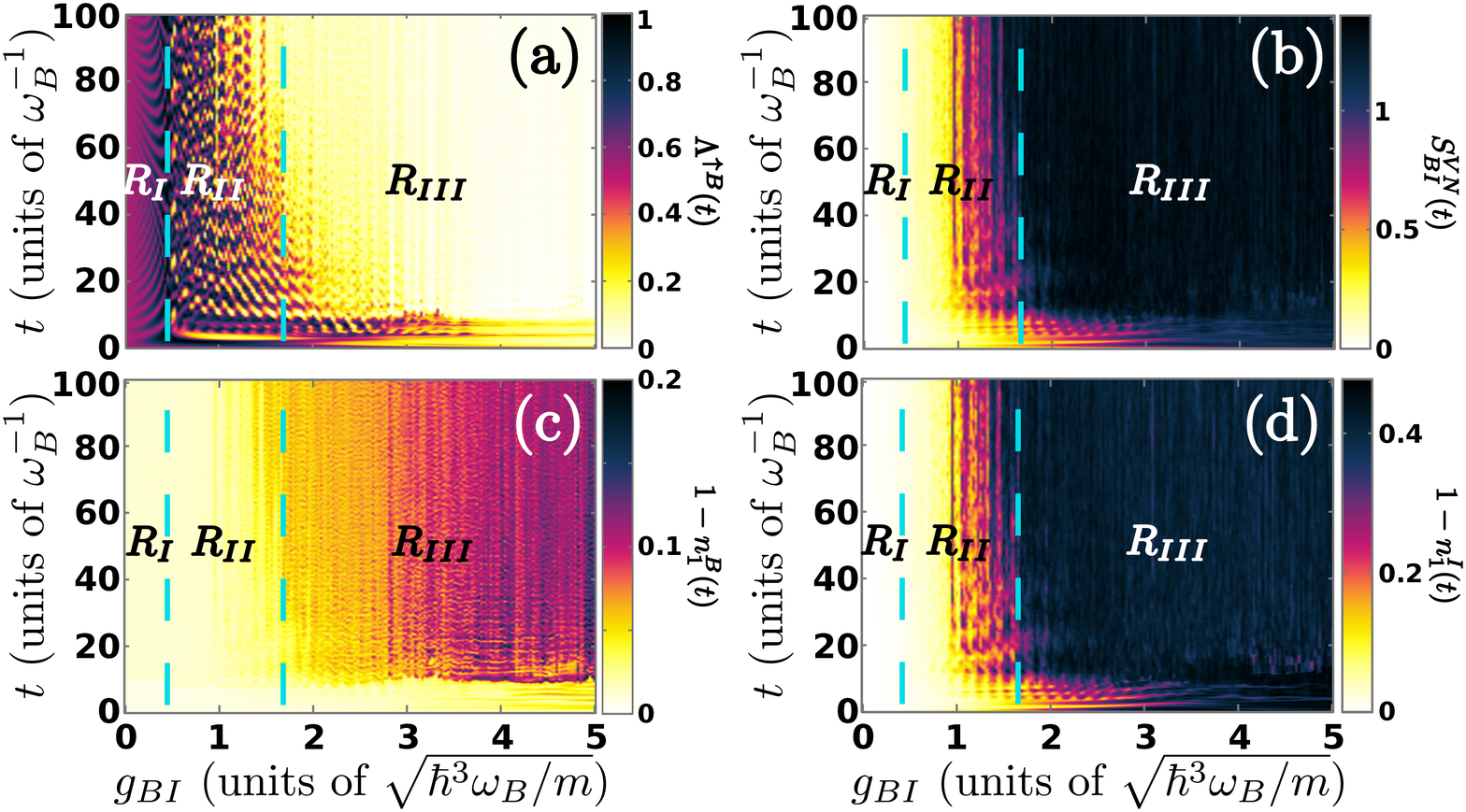}
\caption{(Color online) (a) Temporal evolution of the overlap, $\Lambda^{B \uparrow}(t)$, between the bath and 
the spin-$\uparrow$ of the impurity atom
(b) Evolution of the von-Neumann entropy, $S_{BI}^{VN}(t)$ for increasing $g_{BI}$. 
Deviation from unity of the first natural orbital of the (c) bath and (d) the impurity for varying $g_{BI}$.
In all cases $N_{B}=100$, $N_I=1$ and $g_{BB}=0.5$.} 
\label{Fig:4}
\end{figure}

To complete the physical picture we next focus on the dynamics between the impurity and the bath. For this purpose we employ the 
overlap integral $\Lambda^{B \uparrow}(t)$ [Fig.~\ref{Fig:4}(a)] between the single-particle densities of the bath and the spin-$\uparrow$. 
As dictated by $\Lambda^{B \uparrow}(t)$ within $R_I$ the impurity is partially miscible with the BEC while it is well separated within $R_{III}$.       
The link between $|S(t)|$ and $\Lambda^{B \uparrow}(t)$ is of significant importance not only due to the experimental relevance of     
both quantities but most importantly because $\Lambda^{B \uparrow}(t)$ can be expressed in terms of the Schmidt 
coefficients, $\lambda_i$, see also Eqs. (\ref{overlap_general}) and (\ref{overlap_weak_entanglement}). 
For instance in the weakly entangled case where $\lambda_1\approx1$ and $\lambda_j\ll1$ 
\begin{eqnarray}
\frac{\Lambda^{B\uparrow}(t)}{\Lambda_0(t)}-1 \propto \sum_{j>1} \frac{\lambda_j(t)}{\lambda_1(t)},
\label{Lbi_ent}
\end{eqnarray}
with $\Lambda_0(t)$ being the overlap integral accounting only for the contribution of the 
first Schmidt coefficient $\lambda_1$. 
Therefore $\Lambda^{B\uparrow}(t)$ can be used as a measure to probe the generation of entanglement 
in the MB system.
However to directly visualize the degree of entanglement during the dynamics we employ the 
von-Neumann entropy, $S^{VN}_{BI}(t)=-\sum_i \lambda_i(t) \log \lambda_i(t)$~\cite{Horodeckix4}. 
The temporal evolution of $S^{VN}_{BI} (t)$ [Fig.~\ref{Fig:4}(b)] shows that the dressed impurity 
is entangled with the BEC within the regions $R_I$ and $R_{II}$. 
Most importantly the system becomes strongly entangled within $R_{III}$, where the polaron ceases to exist,
showcasing a plateau of $S^{VN}_{BI} (t>15)\approx 1.2$ for fixed $g_{BB}=0.5$ and 
for all $g_{BI}\gtrsim 1.65$.  
This result is in turn related to the fragmented nature of the system [Figs.~\ref{Fig:4}(c), \ref{Fig:4}(d)]. 
The latter is captured by the deviation from unity of the first natural orbital $1-n^{\sigma}_1 (t)$ of the $\sigma$-species.
Since we consider a single impurity $n^{I}_1 (t)=\lambda_1(t)$ follows $S^{VN}_{BI}(t)$. 
However, the bath fragmentation is almost zero in $R_I$ and $R_{II}$ and it is weak in $R_{III}$. 
Therefore it is the entanglement between the impurity and the bath that plays the crucial role in the 
quasiparticle formation, e.g. see that the growth rate of $S^{VN}_{BI}$ becomes maximal in $R_{II}$, and 
not the fragmentation of the bath.

{}


\begin{thebibliography}{60}
\bibitem{comment_MF} Within the mean-field approximation for both $N_B=100$ and $N_B=10$ cases, 
$|S(t)|$ remains finite for $0<g_{BI}<5$ and thus the orthogonality catastrophe as captured 
by the contrast is absent.
\end{thebibliography}
\end{document}